# Characterizing the All-Sky Brightness of Satellite Mega-Constellations and the Impact on Astronomy Research

**Harrison Krantz*, Eric C. Pearce*, and Adam Block***

*Steward Observatory, University of Arizona*

## ABSTRACT

Measuring photometric brightness is a common tool for characterizing satellites. However, characterizing satellite mega-constellations and their impact on astronomy research requires a new approach and methodology. A few measurements of singular satellites are not sufficient to fully describe a mega-constellation and assess its impact on modern astronomical systems. Characterizing the brightness and impact of a satellite mega-constellation requires a comprehensive measurement program conducting numerous observations over the entire set of critical variables. Doing so requires not only a complete observing program but also the development of techniques for photometric measurements of low-Earth orbit (LEO) satellites. While photometry of deep space satellites is common, doing the same with LEO satellites presents certain technical challenges.

Utilizing Pomenis, a small-aperture and wide field-of-view astrograph, we developed an automated observing program to measure the photometric brightness of mega-constellation satellites including Starlink and OneWeb. We created software to intelligently schedule daily observations with an optimizing algorithm to observe as many satellites as possible while prioritizing particular satellites and geometries to satisfy project goals. The telescope autonomously executes the observing schedule every twilight capturing up to 60 images of satellites per hour. For each satellite we measure photometric brightness and astrometric position. However, the most valuable insights come from analyzing the ensemble of satellites over a comprehensive range of geometries and phase angles. We characterize the all-sky brightness of satellite mega-constellations through the distribution of measurements and correlate the satellite brightness with time of day, on-sky position, solar geometry etc. We also compare the brightness of different satellite designs and report the efficacy of brightness reduction efforts including Starlink's DarkSat and VisorSat.

We aim to measure the potential impact of mega-constellation satellite obtrusion through an assessment of all-sky brightness: a combination of likelihood of satellite appearance and brightness distributions as a function of sky position and time. Analytical studies done by members of the astronomy community show that the impact on astronomy research is real and potentially very significant, particularly for wide-field surveys like the Rubin Observatory LSST. Our project benefits the astronomy community by characterizing the overall impact of satellite mega-constellations in terms of sky availability and the probability of incidental satellites during planned observations.

# 1. INTRODUCTION

In May 2019 SpaceX launched the first 60 Starlink satellites into Low Earth Orbit (LEO). The train of bright satellites surprised star gazers who noted their pronounced naked eye appearance and astronomers who noted the striped images produced when the satellite trains passed through the field of view (FOV) of their telescopes.

The Starlink project intends to provide low-latency high-speed internet via satellite. To accomplish this, SpaceX is launching numerous satellites into LEO. With regulatory approval for 12,000 satellites and plans for an additional 30,000 satellites, the Starlink mega-constellation is unprecedented in all aspects. Similar, competing projects with their own satellite mega-constellations are in development including OneWeb, Amazon Kuiper and others. With multiple mega-constellations in development there may be as many as 100,000 satellites in the sky by 2030.

Astronomers have tolerated interference from satellites for decades. While considered a nuisance by many, their impacts were uncommon and typically minimal. Faraway satellites in geosynchronous orbits are faint enough that their appearance in images can be removed via software. LEO satellites are typically brighter but their low numbers and visibility during only twilight meant they were novel sights to astronomers. The Earth eclipses LEO satellites during most of the night and so the satellites are only visible during the twilight hours. The new mega-constellations and their large numbers of bright satellites appear more frequently in astronomers' images and often bright enough to saturate detectors, posing significant challenges to mitigate. Bright satellite mega-constellations will likely severely impact wide field surveys like Rubin Observatory's LSST project [1].

In June 2020, The American Astronomical Society (AAS) held the *Satellite Constellations 1* workshop (SATCON1) to discuss the impacts from satellites and possible mitigations. SATCON1 produced multiple recommendations for satellite operators and astronomers, including that LEO satellites should orbit lower than 600 km to inhibit visibility during the darkest hours of night and that apparent brightness should be fainter than $7^{th}$ magnitude to avoid saturating detectors on large telescopes such as Rubin Observatory. However, the most important conclusion is "no combination of mitigations can fully avoid the impacts of satellite trails" [2].

We know satellites will be visible to telescopes but to best inform the astronomy community we need to quantify and predict their impacts so astronomers can design future observing programs and future telescope facilities with considerations for the impact satellites will have on observational astronomy. While there is a long legacy of photometric characterization of deep space satellites, doing the same with fast moving LEO satellites presents a technical challenge. The challenge is compounded by our need to photometrically measure a large number of satellites over a wide range of geometries, lighting conditions, and operational phases. In this paper we summarize our observing campaign of LEO mega-constellation satellites and report on their brightness.

## 2. SATELLITE CONSTELLATIONS

There are currently two active satellite mega-constellations on orbit: Starlink and OneWeb. We maintain lists of both constellation member satellites in our targets list for observation.

### 2.1    STARLINK

SpaceX's Starlink satellite constellation consists of 1622 members on orbit as of August 2021. An additional 118 satellites were also launched but have since deorbited [3][4].

Starlink satellites feature a simple L-shaped design (see *Figure 1*) with a thin rectangular bus and a large solar array at a perpendicular angle. The rectangular satellite bus is approximately 3.0 m x 1.5 m and constructed of primarily machined aluminum with many structural features such as thrusters and nadir facing planar antennae. The majority of Starlink satellites operate in a circular 550 km orbit with a 53.0º inclination. A small number of satellites operate in a circular 560 km orbit with a 97.5º inclination. The upcoming second shell of satellites will operate in a circular 570 km orbit with a 70.0º inclination.

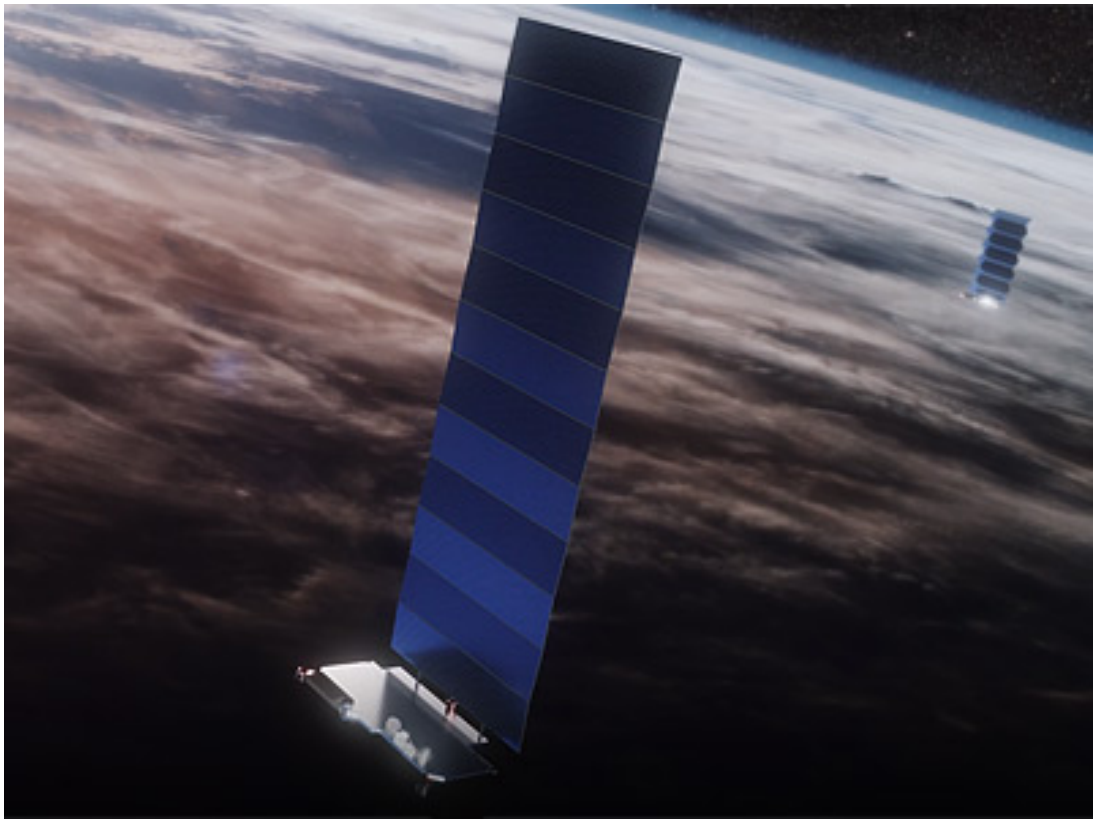

***Figure 1.*** *A rendering of a Starlink satellite with solar array extended. Image courtesy SpaceX.*

#### 2.1.1    STARLINK VISORSAT

To reduce the apparent brightness of the Starlink satellites, SpaceX added a sun visor to the satellite bus (see *Figure 2*). This visor hangs below the satellite and blocks sunlight from hitting the nadir facing surface and reduces the amount of light reflected to observers on Earth. SpaceX launched a prototype of this design, dubbed *VisorSat*, on June 4th 2020. Subsequently all new Starlink satellites feature the modified design starting on August 7th 2020 [5].

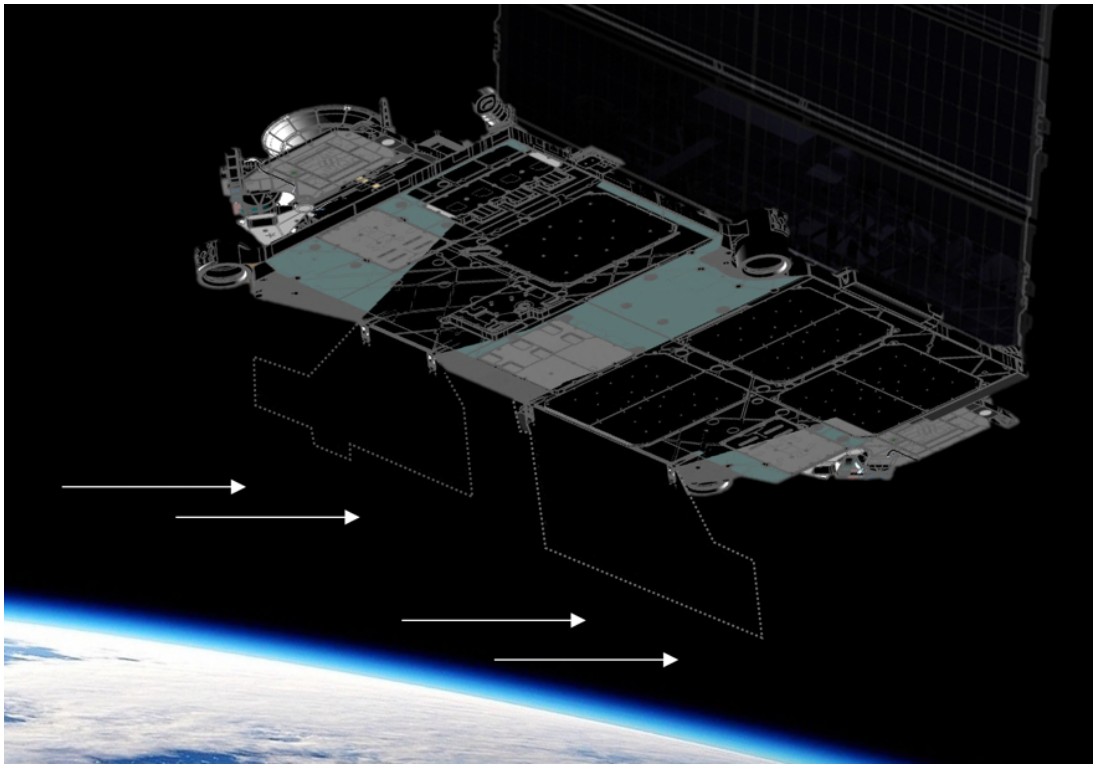

***Figure 2.*** *A rendering of the Sun visors shading the underside of the Starlink satellite bus. Image courtesy SpaceX.*

### 2.1.2 STARLINK DARKSAT

Before implementing the *VisorSat* design SpaceX earlier tested a prototype satellite with a darkened coating dubbed *DarkSat*. SpaceX launched *DarkSat* on January 7th 2020. *DarkSat* featured a dark coating on the nadir facing antennae panels which were previously white (see *Figure 3*). The dark coating reduces the reflected light visible to observers on Earth [5].

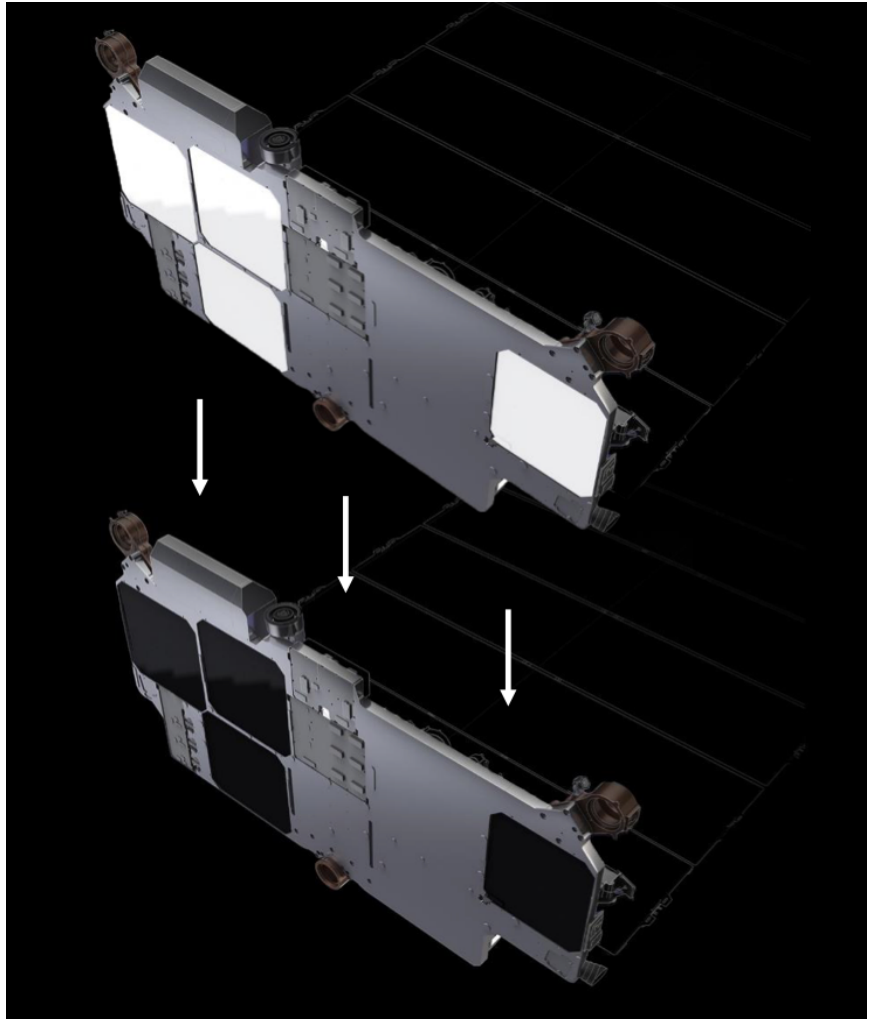

***Figure 3.*** *A rendering of a typical Starlink satellite with white antennae (upper) and the modified dark antennae on DarkSat (lower). Image courtesy SpaceX.*

### 2.2 ONEWEB

The OneWeb satellite constellation consists of 254 members on orbit as of August 2021 [6].

OneWeb satellites feature a typical box-wing design (see *Figure 4*) with a central box-shaped bus and a pair of extended solar panel arrays. Additionally, there are two extended antennae reflectors. The satellite bus is not rectangular but features angled sides and a multi-faceted nadir facing side. The satellite bus also appears to be wrapped in Kapton foil. The satellite bus measures approximately 1.0 x 1.0 x 1.3 m. The original OneWeb constellation design specified a circular 1200 km orbit with an 86.4º inclination. Subsequent modifications to the constellation design may move satellites to a lower orbit.

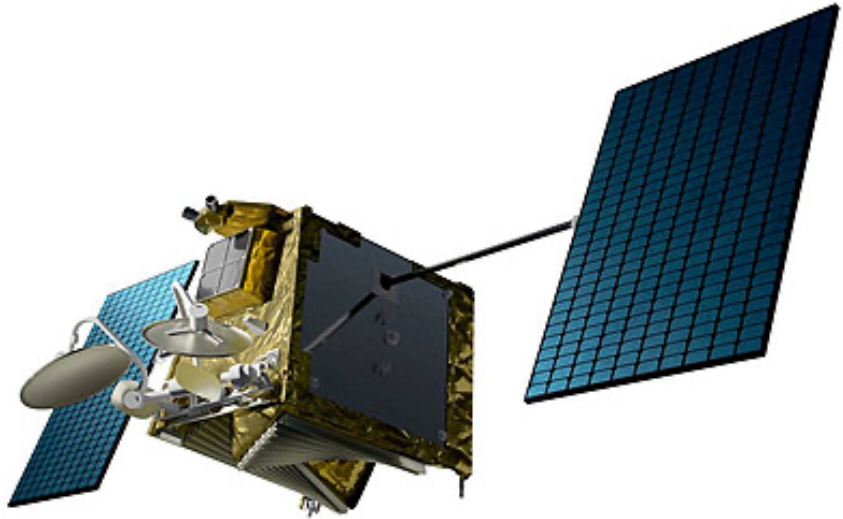

*Figure 4.* *A rendering of a OneWeb satellite showing the box-shaped satellite bus and pair of extended solar panel arrays. Image courtesy OneWeb.*

## 3. METHODOLOGY

Satellite brightness is highly dynamic and dependent on numerous variables including the Sun-satellite-observer geometry. A single measurement does not fully characterize a satellite's brightness, especially in the context of impacts to astronomy observations. The congregate of astronomy studies and observations cover the entire sky warranting characterization of satellite brightness over the entire sky and in all scenarios. To characterize the satellite brightness in all scenarios we devised a comprehensive survey to record measurements across the full range of possible geometries. The survey samples satellites across the entire sky to build a statistical representation of the mega-constellation brightness and visibility. This approach is different from previously reported photometric measurements of singular satellites [1][2][7][8] or with serendipitous observations [9].

We intend to map satellite visibility and potential impact with a holistic characterization of satellite brightness across the entire sky. The map will inform astronomers when and where satellites are visible and the potential impact on planned observations as a function of brightness and apparent movement. With an all-sky map, astronomers will be able to optimize observation scheduling to minimize impact from satellites.

### 3.1 INSTRUMENTATION

For our observing we utilize the Pomenis Observatory [10]. Pomenis is a unique system we specifically created to perform various observations of SSA targets. The telescope itself is a 180 mm Takahashi astrograph which provides a 4.2 x 4.2 degree FOV on a 3056 x 3056 CCD imager with 7-color filter wheel. The telescope rides on a Paramount MYT German equatorial mount.

The system is fully robotic and automated, allowing for remote operation and intelligent automated observing. We utilize the *ACP Observatory Control* software to manage automated system operation [11]. The system includes a unique portable trailer-mounted enclosure. The mobile enclosure enables relocation for different observing programs with minimal necessary site infrastructure. For the majority of observations reported in this paper, Pomenis was located at the summit of Mt Lemmon near Tucson AZ. Pomenis was temporarily relocated to the Biosphere 2 from August to mid-October 2020 due to the Bighorn wildfire which forced closure and evacuations from Mt Lemmon.

The exceptionally large FOV and robotic operation of Pomenis make it particularly capable of observing fast-moving satellites. Observing LEO satellites requires precise timing of the telescope system; the telescope needs to point at the correct spot in the sky and trigger the camera at the exact moment the satellite is flying through the FOV. The robotic operation makes this execution seamless and enables the observation of many satellites in quick succession. The large FOV allows for a larger tolerance of error in timing and error in the actual satellite position. The large FOV also captures the entire satellite streak in the image frame simplifying the photometry and reducing possible error versus measuring the flux density of a truncated streak and estimating the satellite velocity for an effective exposure time. Capturing the entire satellite streak also allows for unambiguous astrometric position determination.

The small telescope size and high-performance mount enable Pomenis to track the fast-moving LEO satellites as well. The pointing error during tracking is often significantly higher but is not an issue with the FOV which is large enough to still capture the satellite.

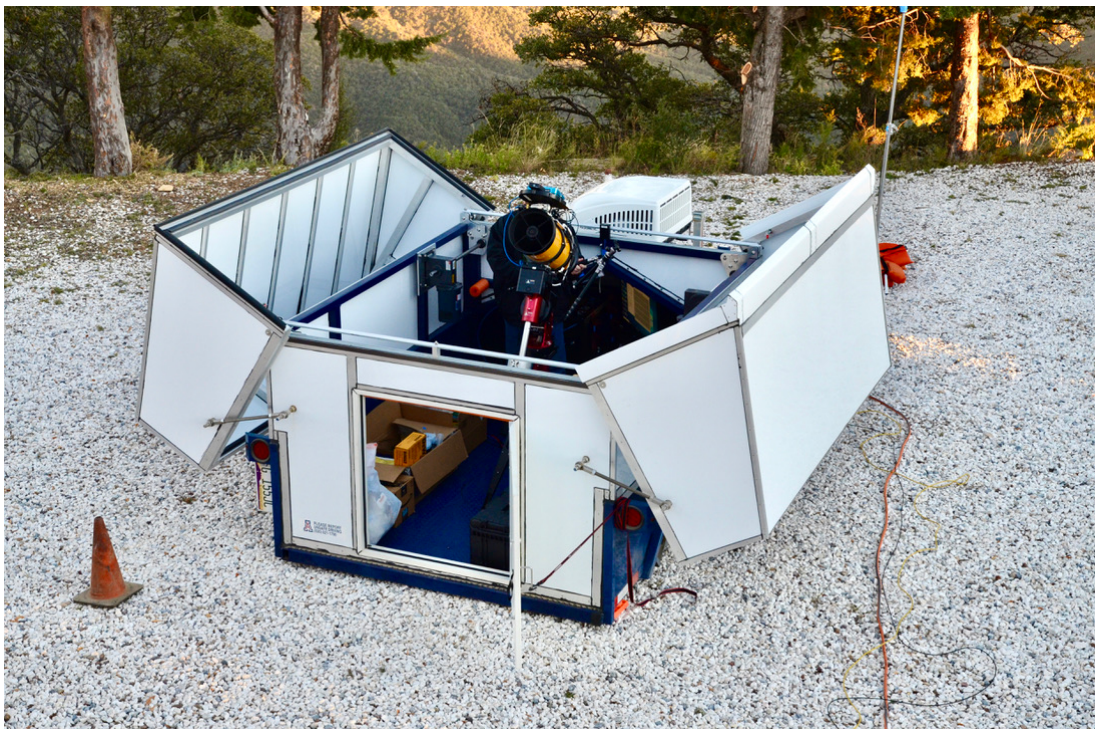

***Figure 5.*** *Pomenis and the unique portable trailer-mounted enclosure.*

## 3.2 OBSERVATION PLANNING AND EXECUTION

We created a custom Python program to plan observations. Utilizing the *SkyField* library [12], the software program calculates all visible satellites from our targets list for the observing session: evening or morning. A satellite is visible if it is above the horizon limit and illuminated by the Sun. The software program then selects which satellites to observe with an optimizing algorithm to observe as many satellites as possible while prioritizing particular satellites to satisfy project goals. Currently the software program is configured to schedule the observation at a random point during a given satellites fly-over pass. The random sampling yields a statistical representation of the population of satellites. Lastly, the software program writes the schedule as an *ACP Plan* for observation by the telescope.

During the survey, the telescope observes each satellite while tracking the background stars at a sidereal rate. This results in an image with a satellite streaked across it as shown in *Figure 6*. Due to the exceptionally large FOV of the Pomenis telescope, the satellite takes multiple seconds to cross the FOV. Thus we are able to open and close the shutter and produce a finite streak wholly contained within the image. This allows for simple photometry by summing the flux in the streak and comparing to background stars. Through experimentation we decided on 3.0 second exposures as this yields a balance of sufficient flux from background stars and high likelihood of the entire satellite streak being within the image frame. Additionally, we capture a clean background image of each star field before the satellite's arrival. We utilize this clean image to subtract any background stars from the imaged satellite streak during processing.

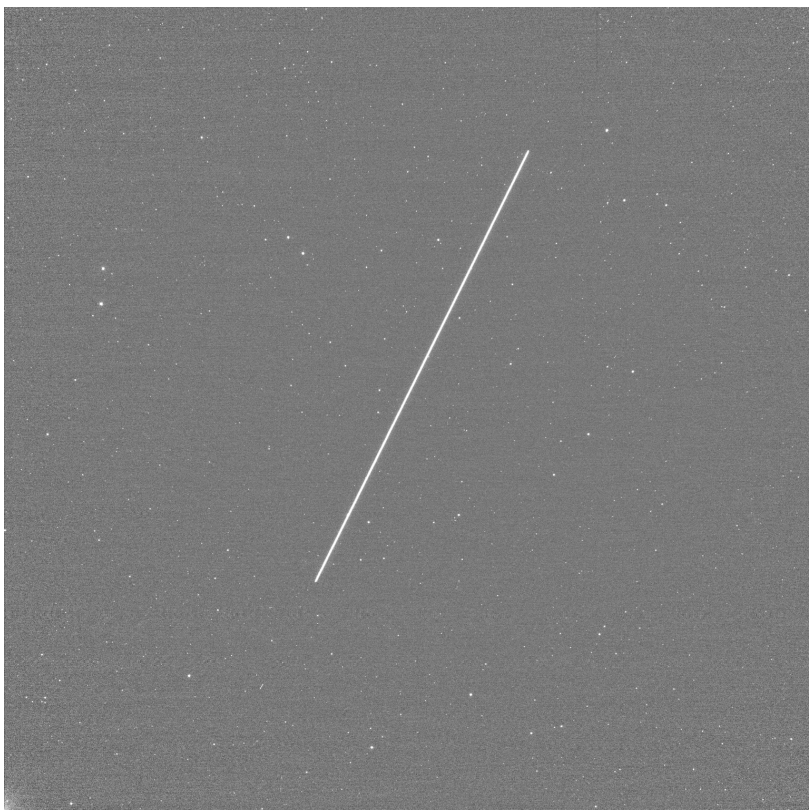

***Figure 6.*** *An image of a bright satellite streak recorded by Pomenis.*

### 3.2.1 TRACKING SATELLITES

To supplement our survey observations, we conduct tracked observations of specific satellites. Whereas the survey observations sample many satellites over a wide range of geometries, the tracked observations produce a high-fidelity lightcurve of the satellite's brightness and variations during a single fly over pass.

We track the satellites utilizing the publicly available Two Line Element set (TLE). During the observation the imager records consecutive images of 3.0 seconds each. To reduce the CCD readout time and overhead between images we bin the detector 2x2 which yields an effective rate of 1 image per 4.7 seconds. We begin tracking the satellite as soon as it is above horizon obstructions and continue until it reaches the opposite horizon or enters eclipse. If the satellite crosses the meridian there is a break in measurements while the telescope completes a meridian flip and reacquires the satellite.

Due to the increased time per satellite compared to the survey observations, fewer satellites can be imaged per night. However, the additional measurements and lightcurves yield interesting insights into the dynamic brightness of the satellite because the viewing geometry and phase angle continuously varies during the pass.

As part of a separate project, Pomenis conducts all-sky night sky brightness survey measurements which we exploit to build extinction models and use to calibrate our tracked satellite observations where differential photometry is not possible due to severely streaked background stars. By eliminating interruptions for observing reference stars, the adventitious use of the all-sky calibration data significantly improves the efficiency of our observations.

### 3.3 IMAGE PROCESSING

Traditional photometry tools are not equipped to process images of streaked satellites. We built our own custom software to address the unique requirements of our project, including a semi-automated pipeline to efficiently process the thousands of images. Our software is primarily *Python* based and makes extensive use of the *Astropy* library [13][14].

The images are reduced via standard practice of bias subtraction and flat field division. After reduction, the software automatically extracts star positions from the image and queries *Astrometry.net* [15] for an astrometric solution which is added to the image metadata in the form of a World Coordinate System (WCS).

The unique challenge in measuring the brightness of fast-moving LEO satellites is performing photometry on a large streak. As described in section 3.2, we purposefully time the Pomenis image acquisition such that the entire streak is within the image frame. This simplifies the photometry to summing the total flux of the streak. We tested various algorithms and methods for automatically detecting the streak within the image but found that all methods resulted in detection failures which ultimately require manual review by a human for data quality assurance. In particular, detection failed for faint satellite streaks which if uncorrected would unreasonably bias the dataset to only bright satellites. Thus we decided to create a software widget for effective manual selection of the streaks in each image.

To determine the photometric zero point for each observation we utilized the *Photometry Pipeline* [16]. This software program is intended for measurements of asteroids and minor planets, though we have utilized a modified version for our other observing programs with Pomenis.

The photometric measurements of all targets and associated metadata including measured astrometric position are compiled in a *MySQL* database for organization and easy query. During post processing we pair each measurement with the contemporaneous TLE and compute the ephemeris and geometry for each measurement.

## 4. RESULTS & ANALYSIS

In total we have 7631 successful photometric measurements as of June 2021. Measurements which we initially rejected may be recovered with additional processing and further measurements may be rejected after audit of the dataset. Unless otherwise noted, the brightness measurements reported are as-observed, only corrected for airmass extinction as is the result of differential photometry. Ultimately astronomers whose observations are impacted by satellites are seeing the satellites as they are in the sky. Normalizing all the measurements to a standard range at zenith can be useful for direct comparisons but does not accurately represent observed brightness of satellites as they impact astronomical observations. Additionally normalizing the measurements to a standard position is erroneous without robust consideration of the peculiar effects that geometry imparts on the satellite brightness.

### 4.1 GENERAL SUMMARY

The mean brightness and general statistics for each population of satellites are listed in *Table 1*. The corresponding distributions are shown in
*Figure 7*.

| Constellation | n Measurements | Population Mean V-mag | Population Std Dev | Brightest Observed |
|---|---|---|---|---|
| **Standard Starlink** | 3146 | 7.0 | 1.1 | 1.5 |
| **Visored Starlink** | 2263 | 8.0 | 1.1 | 2.7 |
| **OneWeb** | 1456 | 9.1 | 1.0 | 5.4 |

*Table 1. General statistics for the as-observed brightness of each satellite population.*

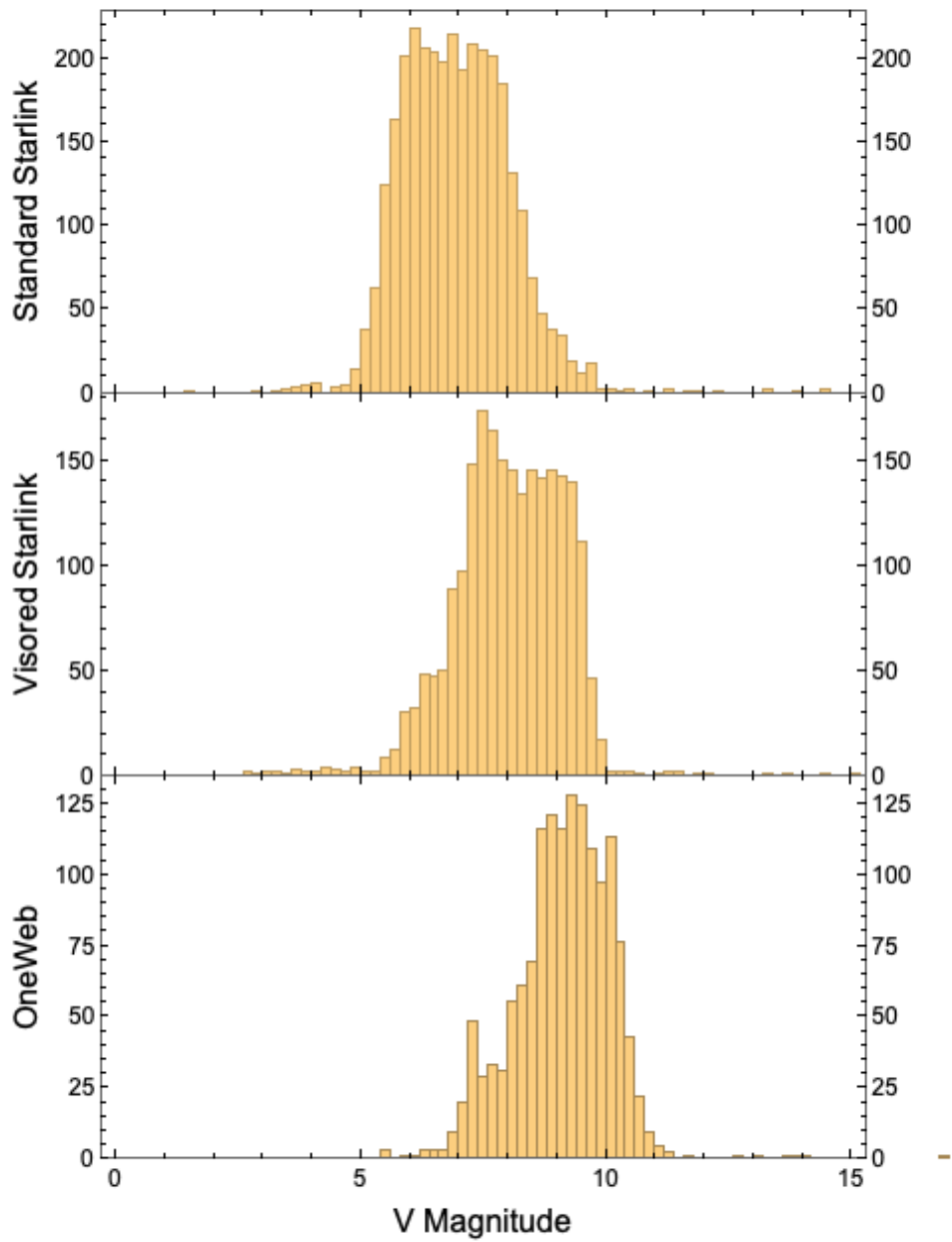

***Figure 7.*** *Histograms of the as-observed brightness for standard Starlink, visored Starlink, and OneWeb satellites. The distribution shows the typical range of apparent brightness.*

The distribution of brightness measurements represents the typical brightness of a randomly observed satellite somewhere in the sky. The standard Starlink satellites exhibit a typical brightness of ~5th magnitude to ~9th magnitude while the visored Starlink satellites are dimmer at

~6th magnitude to ~10th magnitude. At ~7th to 11th magnitude the OneWeb satellites are dimmer than both populations of Starlink satellites.[1]

Every population exhibits a large range of typical brightness warranting a large survey effort to capture the full range. Individual satellite brightness is highly dynamic and dependent on numerous variables including the observer-satellite-Sun geometry, satellite attitude and pose, and satellite structural design.

### 4.2 RELATION TO GEOMETRIC PARAMETERS

Here we present the as-observed brightness plotted against common parameters for modeling satellite brightness. Modeling satellite brightness is complex because the key parameters which determine apparent brightness are not independent variables. A correlation seen with one parameter is likely intertwined with multiple other parameters.

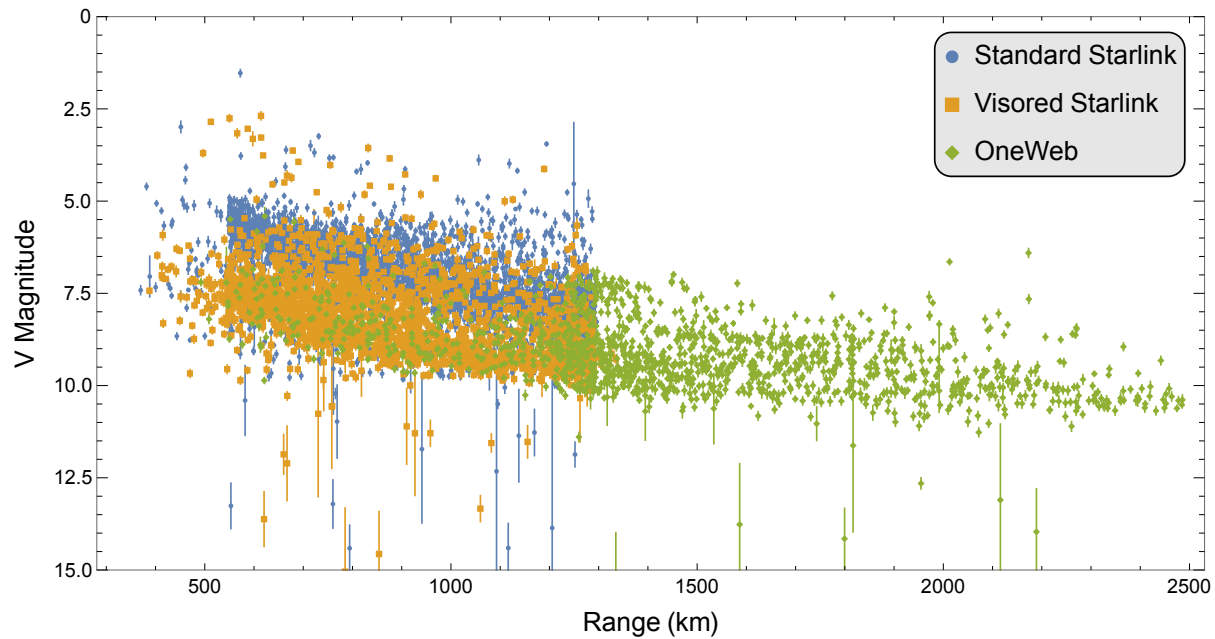

*Figure 8. The satellite brightness shows an expected correlation with range with farther satellites appearing dimmer in the sky.*

---

[1] *N.B. The average brightness reported here is dimmer than what we previously reported in 2020 [2]. This is due to an increased number of measurements low in elevation. Satellites low in elevation have an increased range and are typically dimmer. Furthermore more satellites are visible low in elevation, thus a randomly sampled population will have a bias to dimmer, lower elevation satellites. We only sampled satellites high in the sky during our initial summer 2020 observing survey which was then changed to sample randomly over the entire sky.*

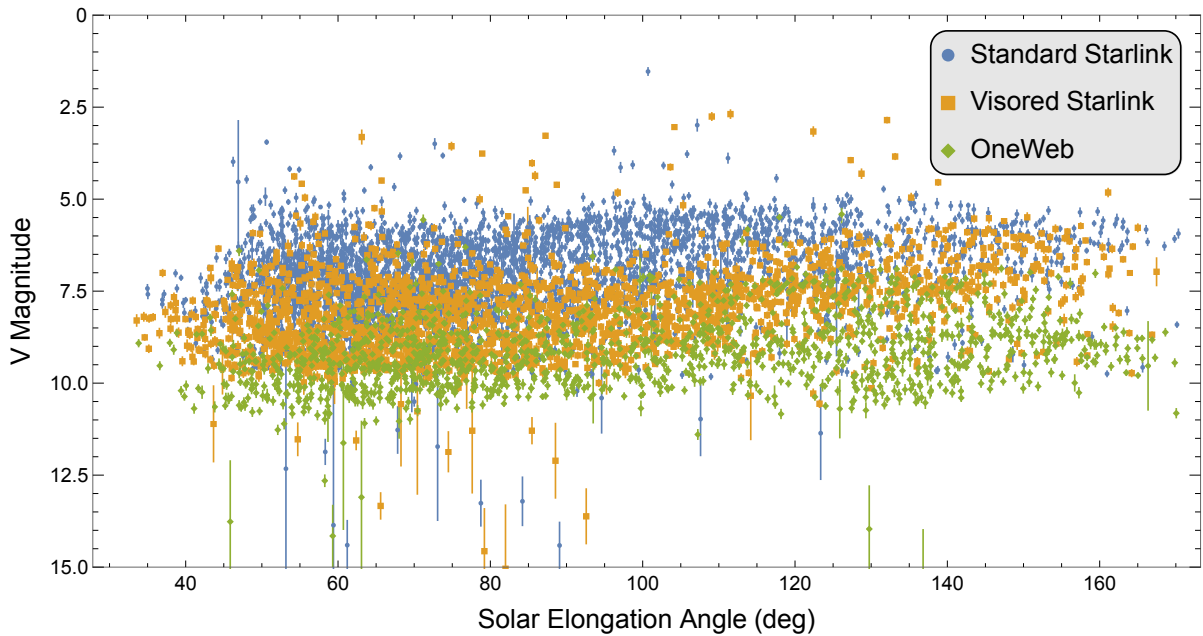

*Figure 9.* Satellite brightness does not show a clear correlation with solar elongation. Solar elongation is defined as the Sun-observer-satellite angle; i.e. the apparent angular separation between the Sun and satellite.

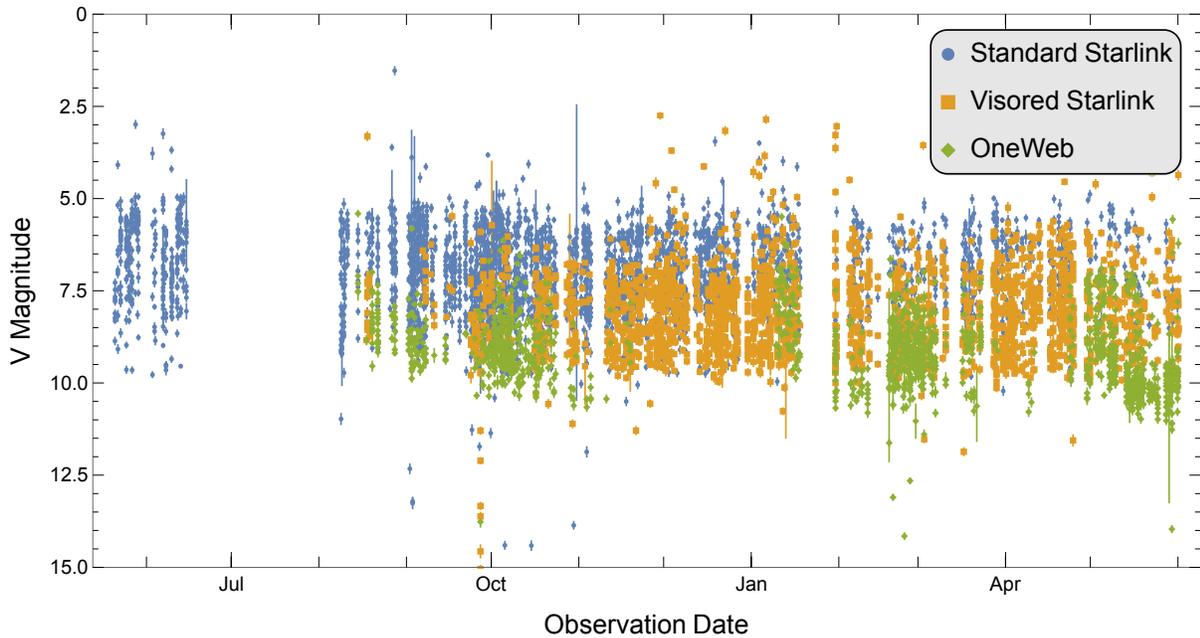

*Figure 10.* A year of satellite brightness measurements plotted over observation date from May 2020 to June 2021. Our initial observing survey of only Starlink satellites ran in May & June 2020 before shutting down due to wildfires. In August 2020 we restarted the survey with OneWeb satellites added. Visored Starlink satellites also began launching to August 2020. Note that the incomplete OneWeb constellation is observed irregularly due to seasonal visibility blackouts.

## 4.3   EFFECTIVE ALBEDO MODEL

The simplest satellite brightness models simulate satellites as gray diffuse reflecting spheres [17]. To calculate the expected satellite brightness one supplies the satellite's range, phase angle, size, and albedo. The relatively simple formula calculates the amount of light reflected by the satellite and visible to an observer. These simple models do not accurately predict satellite brightness in many scenarios because satellites are not diffuse spheres but structured objects with many specular features. More complicated models attempt to model these specular reflections but often fall short of being universally useful.

Although a simple diffuse sphere model does not consider the structural features of the satellite and specular reflections, it does allow us to create a metric for orientation specific brightness as compared to the diffuse sphere model. We measured the real observed brightness and calculated the geometric variables from the ephemeris, leaving albedo as the only unknown[2]. We reconfigured the diffuse sphere model to calculate the albedo given the other values. If the diffuse sphere model worked in all scenarios we would expect the calculated albedo to be the same for all measurements. Deviations of the calculated albedo indicate when the satellite's brightness matches the model and when it doesn't, i.e. when the satellite's brightness is diffuse reflection dominated vs specular reflection dominated. We call this metric the effective albedo since this calculated albedo is not the real satellite albedo, but rather a measure of the reflection's specularity.

Below we present plots of the calculated effective albedo plotted against various parameters.

---

[2] *We estimate the satellite size from images and published dimensions.*

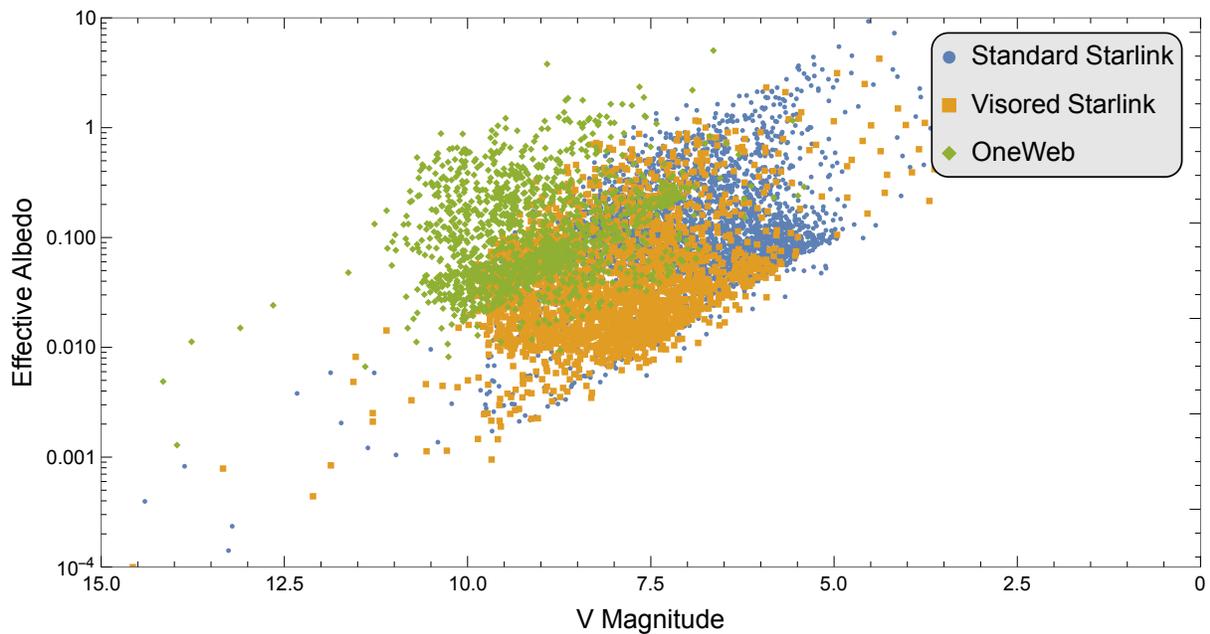

*Figure 11.* As expected, brighter measurements correspond to higher effective albedo. The typical effective albedo is on order of ~0.1 which matches well with typical real satellite albedo. Higher effective albedo indicates the satellite brightness is dominated by specular reflection and no longer follows the diffuse sphere model.

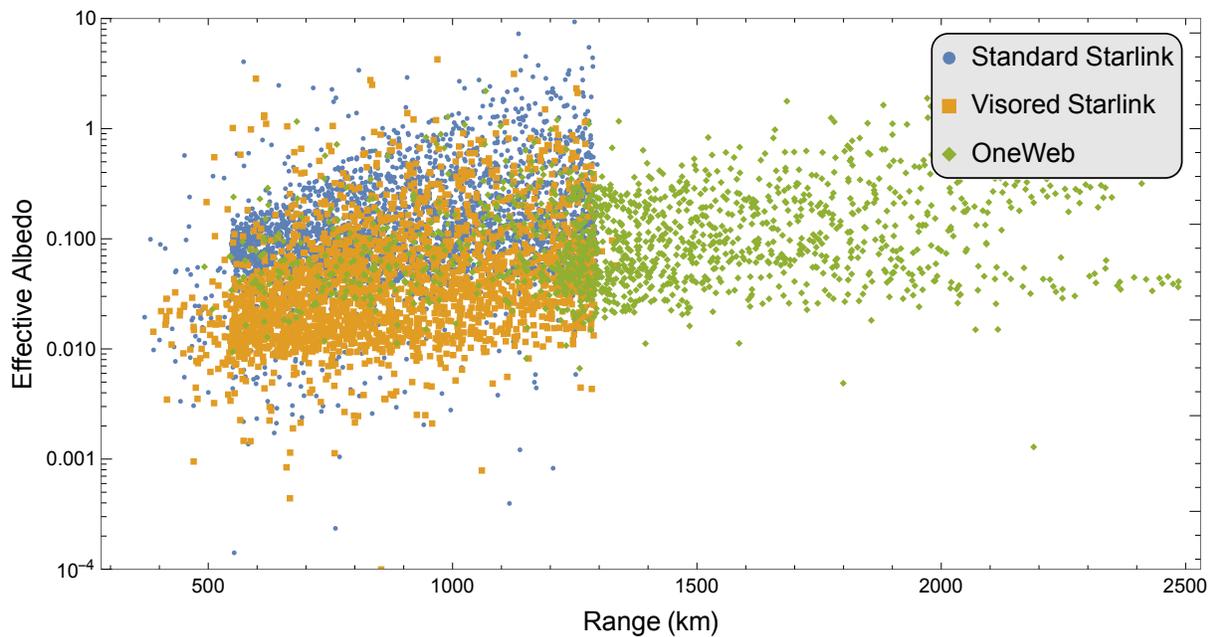

*Figure 12.* The effective albedo does not show a strong correlation with range like the as-observed brightness did in *Figure 8*. Interestingly, each distribution shows a vague trapezoidal arrangement with a sparse number of high effective albedo measurements at larger ranges. This is because a bright specular reflection at far range corresponds to a very high effective albedo.

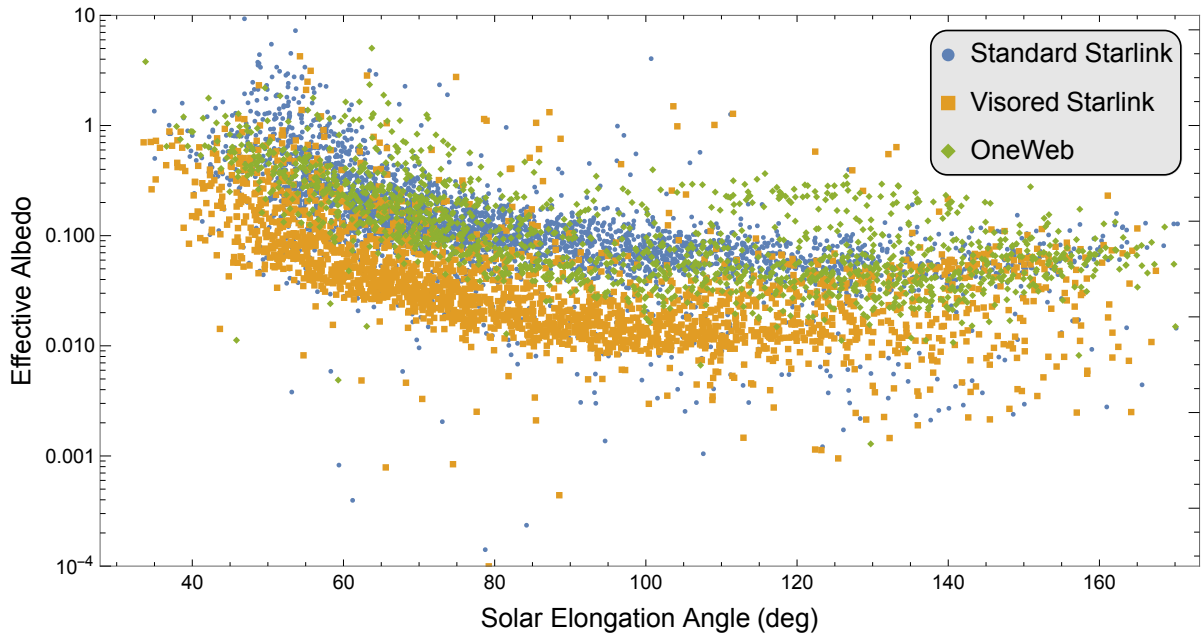

***Figure 13.*** *The most interesting correlation with effective albedo is with solar elongation. Contrary to expectation, the highest effective albedo is at low elongation angles. This is due to sunlight reflecting off the nadir facing surface of the satellite bus at very low glancing angles.*

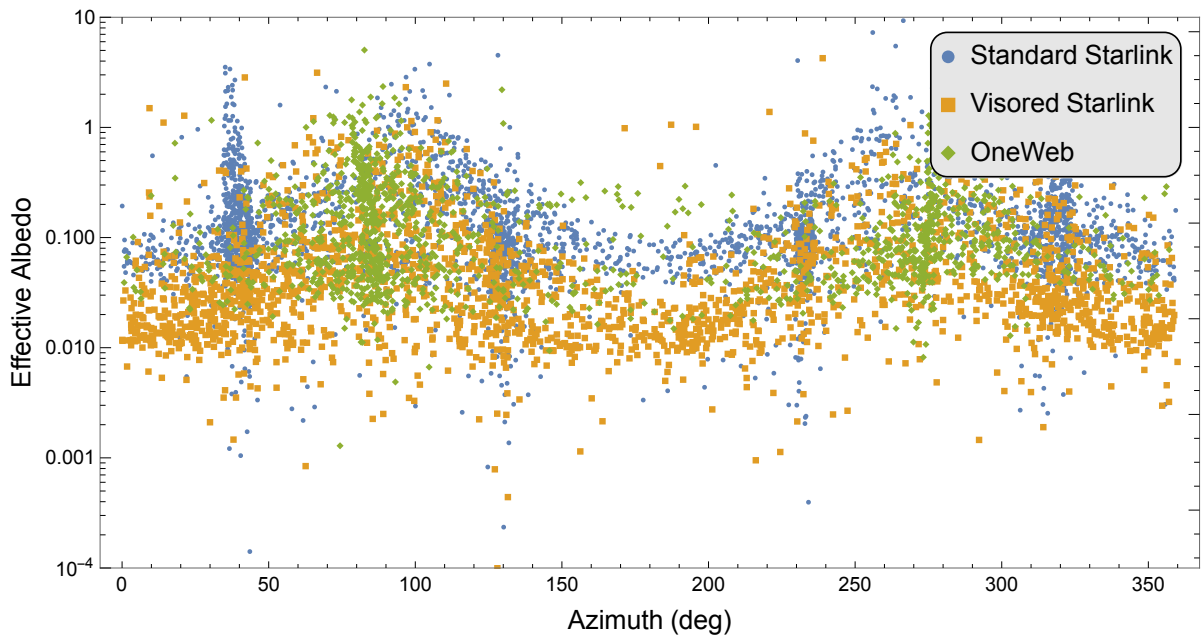

***Figure 14.*** *There is a clear increase in effective albedo at specific azimuth angles which correspond to the Sun's position below the horizon (in the East during morning observations and in the West for evening). This effect is strongly coupled with the specular reflections correlated with low solar elongation. The dense clustering of standard Starlink satellites is a residual effect from the initial summer 2020 observations and the non-random sampling. The dense clustering of OneWeb satellites is because of their high inclination orbit.*

## 4.4 HIGH-FIDELITY TRACKED MEASUREMENTS

Below we present examples of high-fidelity measurements from our tracked observations.

### 4.4.1 ONEWEB

A typical OneWeb track is shown in *Figure 15*. The OneWeb satellites are generally >8th magnitude though consistently increase in brightness near zenith when at closest range. The increase in brightness near zenith is asymmetric with brighter measurements when opposite the Sun, a manifestation of the opposition surge effect. The lightcurve shown in *Figure 16* reveals many discrete flares in brightness which are likely due to the irregular and multi-faceted surface of the satellite bus. The effective albedo plotted in
*Figure 17* shows that the OneWeb satellites are predominantly diffusely reflective with little variation in the effective albedo.

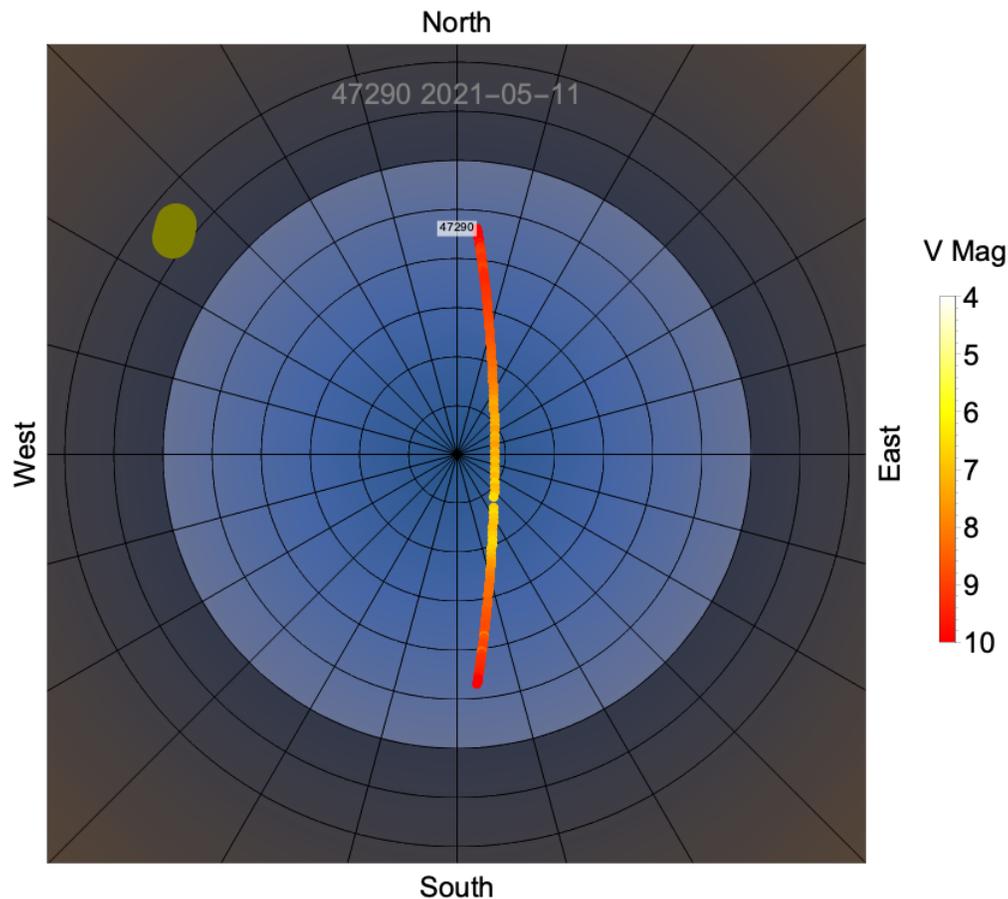

*Figure 15. The brightness of a OneWeb satellite as it passed overhead showing an increase in brightness near zenith.*

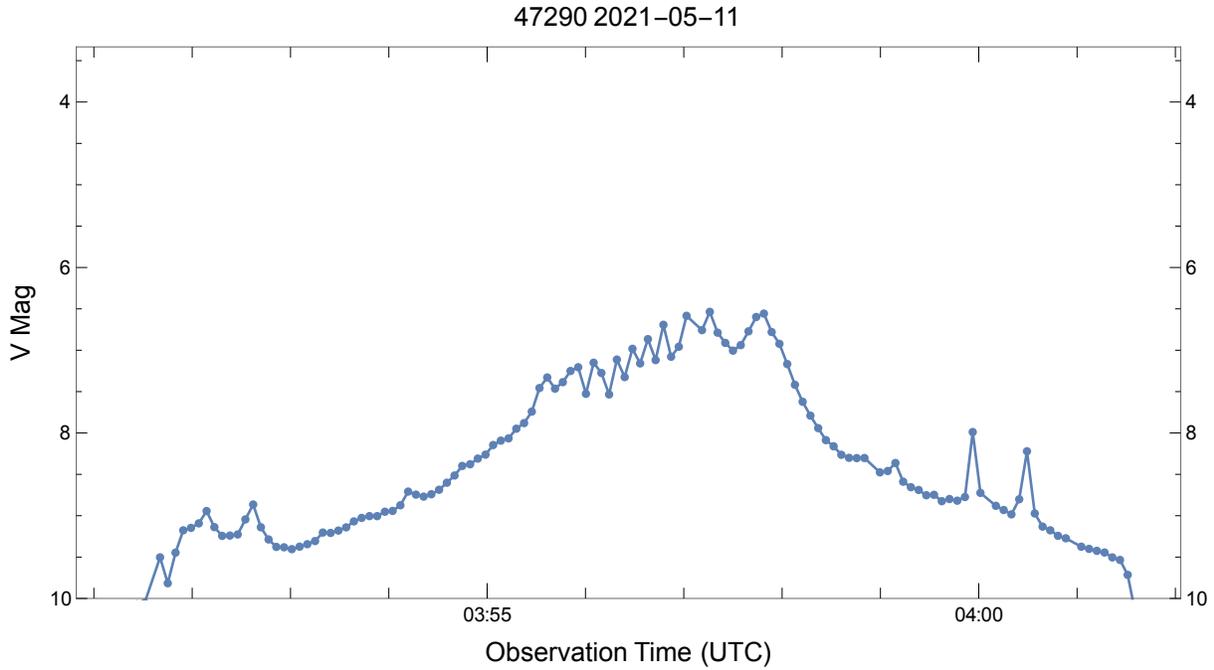

*Figure 16.* *The lightcurve of a OneWeb satellite as it passed overhead. There are many flares in brightness which are likely due to the irregular and multi-faceted surface of the satellite bus.*

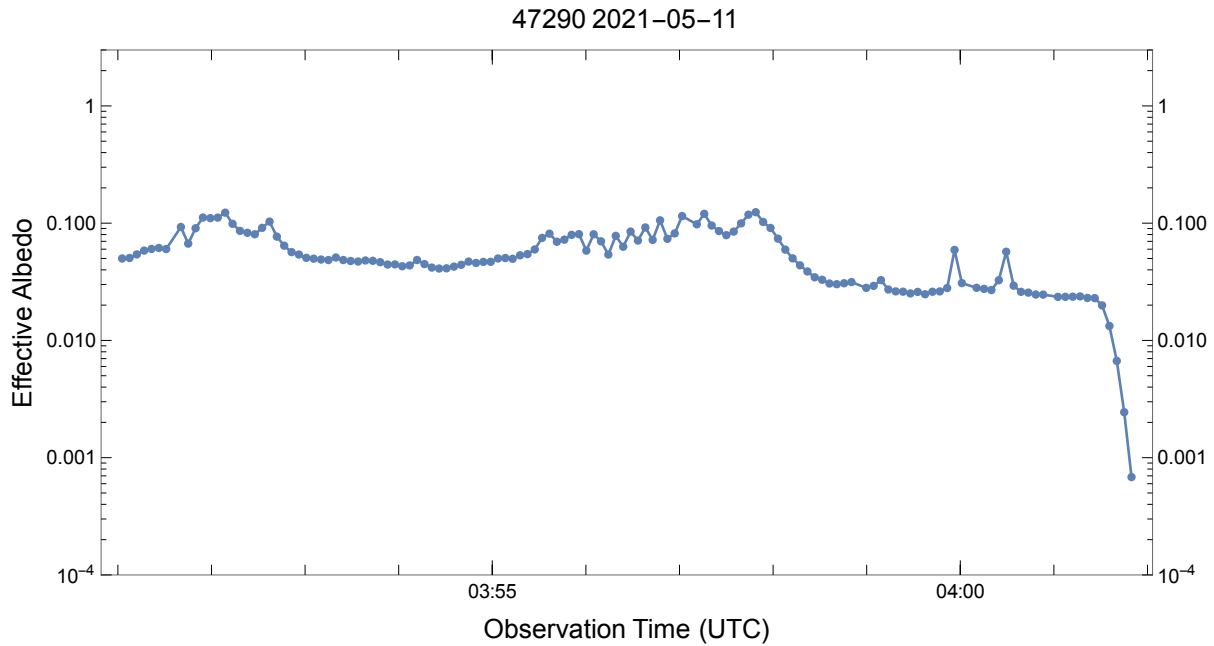

*Figure 17.* *The effective albedo "lightcurve" of a OneWeb satellite as it passed overhead. Although there are many flares in brightness, the effective albedo is relatively constant indicating the satellite brightness was dominated by diffuse reflection.*

### 4.4.2 STARLINK

Like the OneWeb satellites, the Starlink satellites also increase in brightness and exhibit the same opposition surge effect when opposite the Sun near zenith. However, the Starlink satellites display additional geometry dependent increases in brightness away from zenith.

One characteristic example is shown in *Figure 18* where a visored Starlink satellite flares in brightness low in the sky and with very low solar elongation just above the setting Sun. This ~1.5 magnitude increase in brightness is apparent in the lightcurve in *Figure 19*. The flare is even more evident in the effective albedo plotted in *Figure 20* where the effective albedo is significantly higher than the remainder of the measurements, indicating a strong specular reflection. We see this characteristic behavior from all Starlink satellites, visored or not. We suspect this is a low angle glancing specular reflection off the nadir face of the satellite bus. It is not clear if the attitude of the satellite in this geometry impedes the visor's ability to block sunlight hitting the satellite bus.

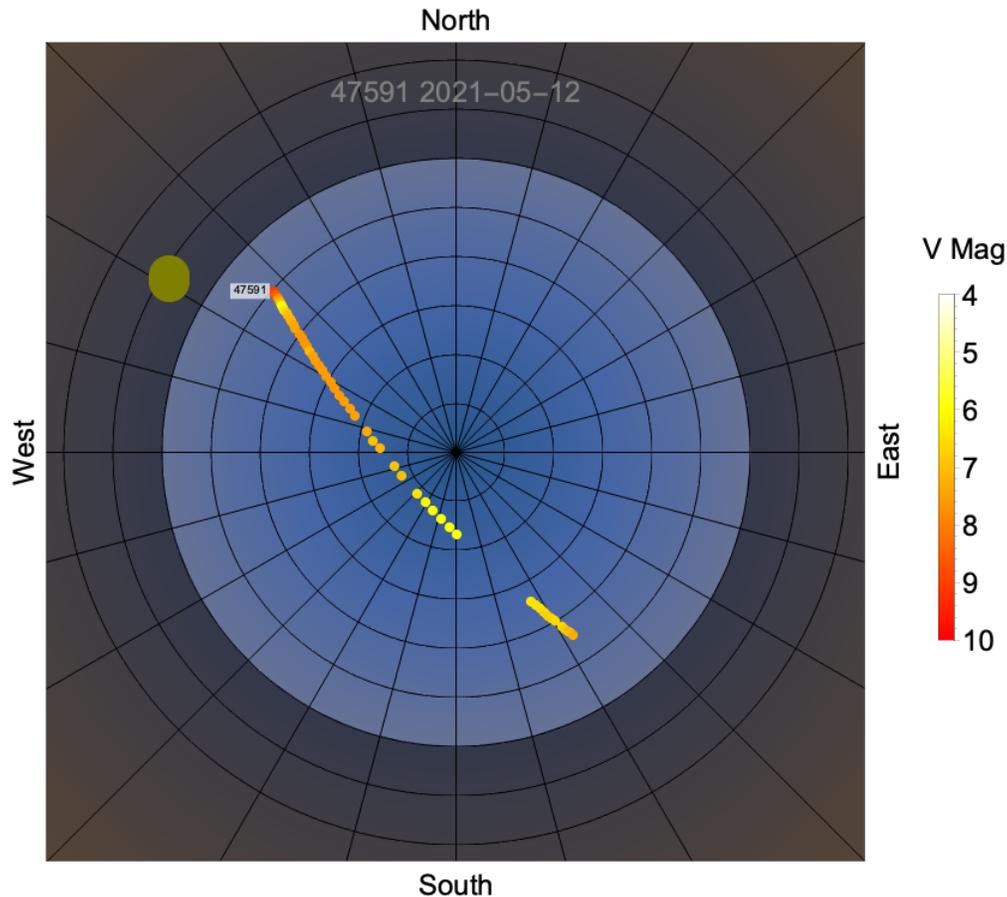

***Figure 18.*** *The brightness of a visored Starlink satellite as it passes overhead. There is a notable increase in brightness near the beginning of the pass near the Sun.*

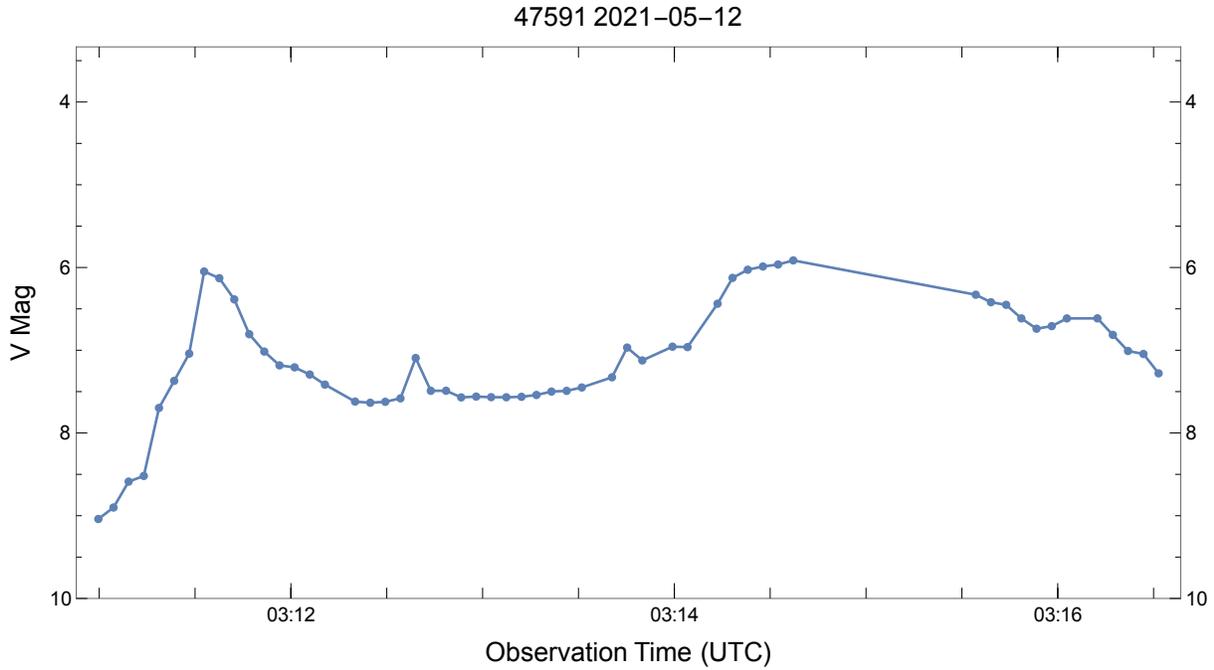

***Figure 19.*** *The brightness lightcurve of a visored Starlink satellite as it passed overhead. There is a ~1.5 magnitude flare in brightness at the beginning of the pass.*

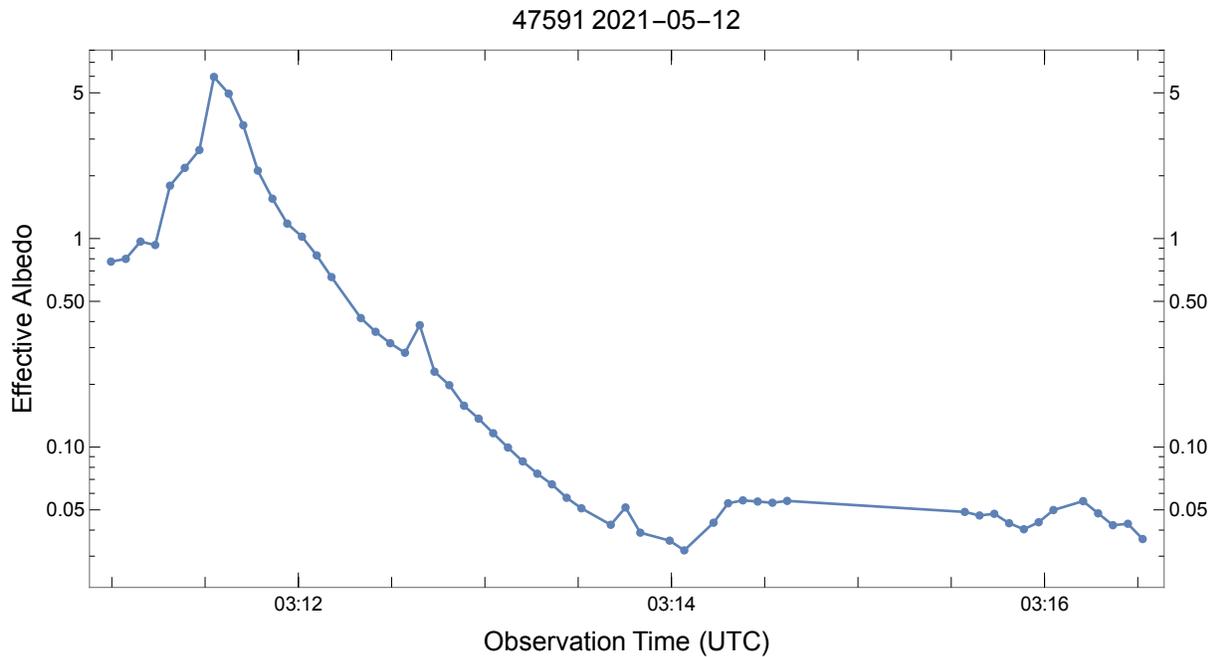

***Figure 20.*** *The effective albedo "lightcurve" of a visored Starlink satellite as it passed overhead. The flare in brightness seen in **Figure 19** corresponds to a massive increase in effective albedo indicating a strong specular reflection at the beginning of the pass at low solar elongation.*

### 4.5 DARKSAT

During the course of our observation survey we recorded 19 measurements of the Starlink *DarkSat* satellite. We also made several tracked observations of *DarkSat*. On average *DarkSat* appears 0.7 magnitudes dimmer than the standard Starlink satellites but not as dim as the visored Starlink satellites or the OneWeb satellites.

| Constellation | n Measurements | Population Mean V-mag | Population Std Dev | Brightest Observed |
|---|---|---|---|---|
| **Starlink DarkSat** | 19 | 7.7 | 1.2 | 5.3 |

*Table 2.* General statistics for the as-observed brightness of the DarkSat satellite

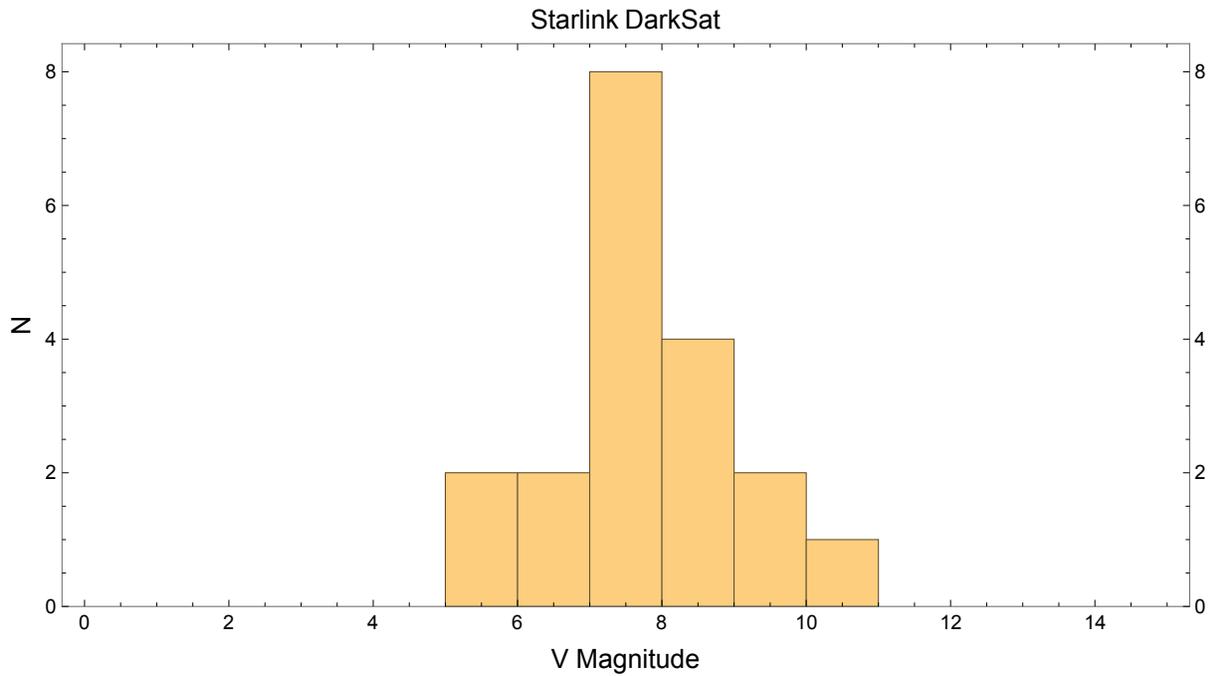

*Figure 21.* Histogram of the as-observed brightness for the Starlink DarkSat satellite.

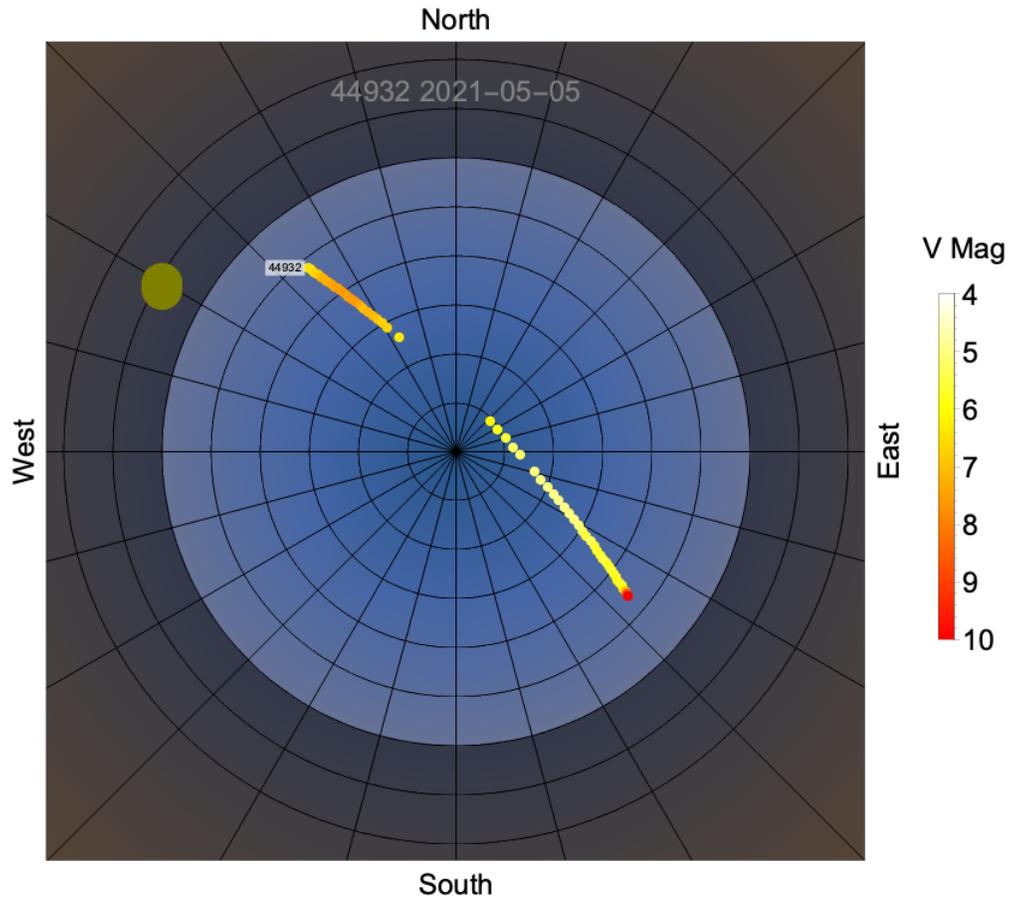

*Figure 22. The brightness of DarkSat (44932) as it passed overhead. Like the other Starlink satellites there is an increase in brightness near zenith, an opposition surge effect, and the characteristic flare at low solar elongation.*

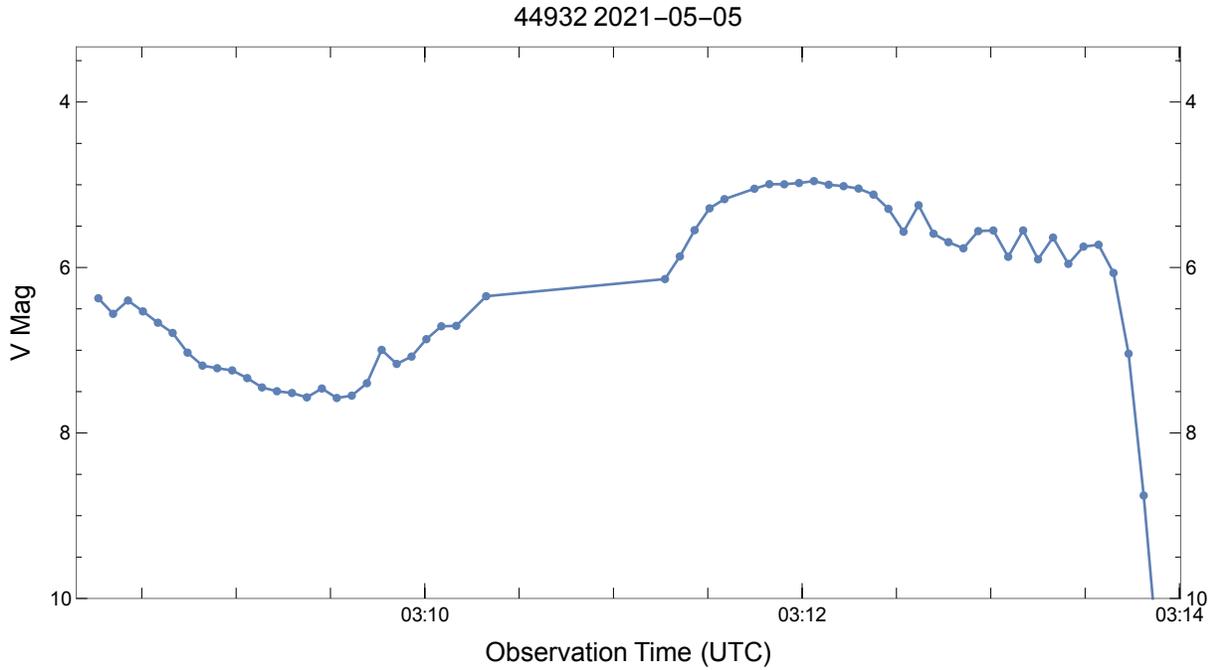

***Figure 23.*** *The brightness lightcurve of DarkSat (44932) as it passed overhead. The brightness ranges from 5<sup>th</sup> to 8<sup>th</sup> magnitude during the pass with multiple glints and discrete variations in brightness.*

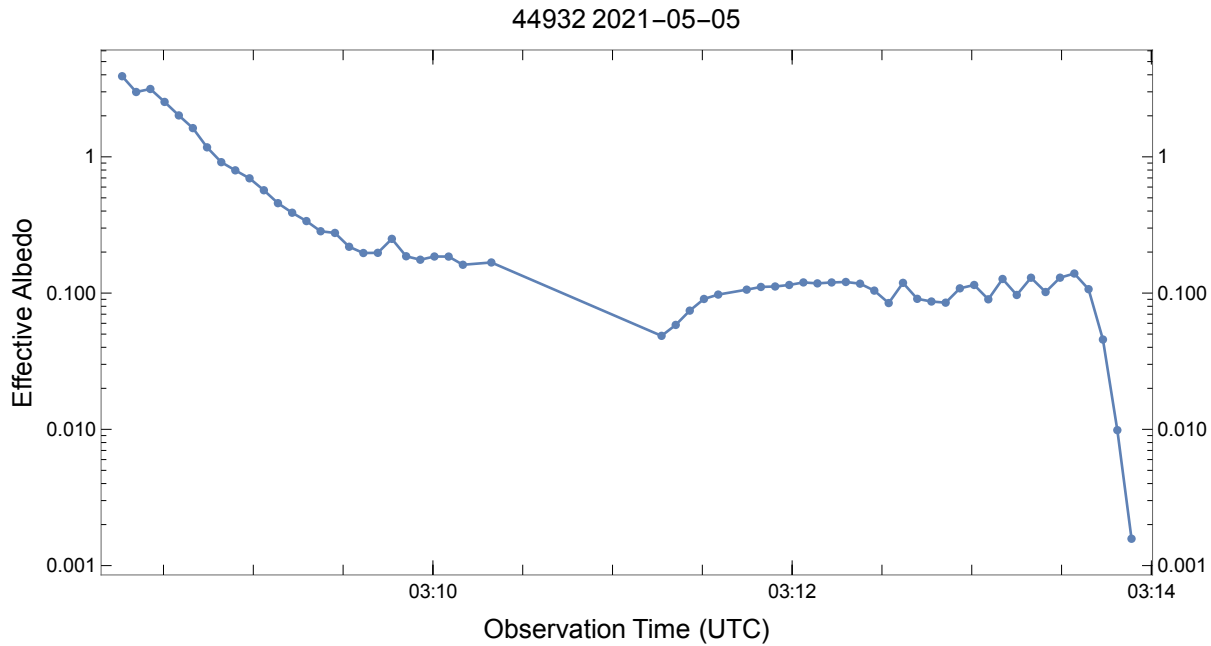

***Figure 24.*** *The effective albedo "lightcurve" of DarkSat (44932) as it passed overhead. The effective albedo is relatively constant around 0.10 during the second half of the pass, but surges to high values during the first half. This is due to specular reflection coming from the nadir facing surface reflecting sunlight at low solar elongation angles.*

## 5. CONCLUSIONS

Starting in May 2020 we began a comprehensive observation survey of mega-constellation satellites in order to characterize their brightness and inform the astronomy community regarding their potential impacts.

With thousands of measurements sampled across the entire sky we report the as-observed brightness distributions of three populations of satellites: standard Starlink, visored Starlink, and OneWeb. The addition of Sun visors to the Starlink satellites reduced the typical brightness by 1.0 magnitude. The OneWeb satellites are typically an additional 1.0 magnitude dimmer than the visored Starlink satellites. This is likely due to their higher orbit and increased range.

| *Constellation* | *Mean V-Magnitude* | *Typical Range* |
|---|---|---|
| **Standard Starlink** | 7.0 | $5^{th}$ to $9^{th}$ magnitude |
| **Visored Starlink** | 8.0 | $6^{th}$ to $10^{th}$ magnitude |
| **OneWeb** | 9.1 | $7^{th}$ to $11^{th}$ magnitude |

Accurately modeling satellite brightness is challenging and the common diffuse sphere model does not consider geometry-based specular reflections. However by reconfiguring the diffuse sphere model we utilize the observed brightness to calculate the effective albedo. This serves as a measure of the specularity of the reflected light.

Calculating the effective albedo from brightness measurements over a variety of scenarios reveal a relatively consistent effective albedo around 0.1 but with notable exceptions where the effective albedo significantly increases to values as high at 10. This indicates a geometry where the satellites produce specular reflections rather than diffuse. In particular, this happens at low solar elongation where the satellites appear just above the Sun which is below the horizon.

With this comprehensive data set we intend to develop a unique model for predicting satellite brightness utilizing the effective albedo calculations. Furthermore, we intend to develop meaningful metrics for determining the potential impact on astronomy observations accounting for satellite brightness, apparent motion, and frequency of visibility.

We intend to make this dataset publicly available in the fall of 2021 after we complete verification of the data. Interested parties can contact the author to be notified when the dataset is available.

## 6. ACKNOWLEDGEMENTS

We thank the members of the Steward Observatory Mountain Operations team for their support and maintenance of the Mt Lemmon Sky Center site.

We thank the staff at Biosphere 2 who allowed our observatory to temporarily reside at their site in the fall of 2020 (and for turning off the lights at night).

We thank Victor Gasho for finding a place a store our observatory on short notice after evacuating Mt Lemmon due to the Bighorn wildfire.

We thank the firefighting crews who worked to protect the Mt Lemmon site from the Bighorn wildfire.